\documentclass[a4paper,5p]{elsarticle}
\usepackage{amssymb}
\usepackage{amsmath}
\usepackage{graphicx}
\usepackage{xspace}
\usepackage{tabularx}
\usepackage{color}

\newcommand{\Nbubble}{$N_{\text{bubble}}$\xspace}
\newcommand{\s}{$\sigma$\xspace}
\newcommand{\h}{$h_{0}$\xspace}
\newcommand{\B}{$\tilde{B}_{\text{max}}$\xspace}
\newcommand{\D}{$D(E)$\xspace}

\newcommand{\e}{$\eta$\xspace}
\newcommand{\Nmax}{$N^{\text{max}}_{\text{bubble}}$\xspace}

\usepackage[draft]{changes}
\definechangesauthor[name = nojoon, color = cyan]{NM}
\definechangesauthor[name = ara, color = brown]{AG}
\definechangesauthor[name = subin, color = green]{SK}

\begin{document}
\setcounter{page}{1}
\title{Neural network-based recognition of multiple nanobubbles in graphene}
\author[1]{Subin Kim\corref{cor1}}
\author[2,3]{Nojoon Myoung\corref{cor1}}
\author[2,3]{Seunghyun Jun}
\author[1]{Ara Go\corref{cor2}}

\affiliation[1]{
	organization={Department of Physics, Chonnam National University},
	city={Gwangju},
	postcode={61186},
	country={Republic of Korea}}

\affiliation[2]{
	organization={Department of Physics Education, Chosun University},
	city={Gwangju},
	postcode={61452},
	country={Republic of Korea}}
\affiliation[3]{
	organization={Institute of Well-Aging Medicare $\&$ Chosun University LAMP Center, Chosun University},
	city={Gwangju},
	postcode={61452},
	country={Republic of Korea}}

\cortext[cor1]{These authors contributed equally to this work.}
\cortext[cor2]{Corresponding author: arago@jnu.ac.kr}
\date{\today}

\begin{abstract}
We present a machine learning method for swiftly identifying nanobubbles in graphene, crucial for understanding electronic transport in graphene-based devices. Nanobubbles cause local strain, impacting graphene's transport properties. Traditional techniques like optical imaging are slow and limited for characterizing multiple nanobubbles. Our approach uses neural networks to analyze graphene's density of states, enabling rapid detection and characterization of nanobubbles from electronic transport data. This method swiftly enumerates nanobubbles and surpasses conventional imaging methods in efficiency and speed. It enhances quality assessment and optimization of graphene nanodevices, marking a significant advance in condensed matter physics and materials science. Our technique offers an efficient solution for probing the interplay between nanoscale features and electronic properties in two-dimensional materials.
\end{abstract}

\maketitle

\section{Introduction}

Structural imperfections, including atomic defects and ripples~\cite{hashimoto2004direct,halbertal2017imaging,fasolino2007intrinsic,paronyan2011formation}, represent diverse factors contributing to disorder in graphene. The formation of graphene nanobubbles, characterized by the elastic deformation of crystal structures, is an intrinsic phenomenon observed during sample transfer onto substrates~\cite{levy2010strain,leconte2017graphene}. Extensive research has explored the impact of nanobubbles on the reduction of carrier mobility~\cite{leconte2017graphene}. As a result, researchers strive to avoid regions prone to nanobubble formation when fabricating graphene devices, aiming to mitigate potential, unidentified influences. Recognizing nanobubbles in a graphene sample is thus pivotal for fabricating high-quality devices~\cite{wang2013one}. Nanobubble location being a local feature necessitates the use of scanning probe techniques that focus on processing local information.

Substantial research has focused on the impact of elastic strain induced by nanobubbles, which create a characteristic, non-uniform profile of pseudomagnetic fields (PMFs) \cite{qi2014pseudomagnetic,levy2010strain,yuan2023compressive}. These PMFs play a critical role in the transport of Dirac fermions through graphene, facilitating the emergence of phenomena such as zero-field Landau levels \cite{levy2010strain,nigge2019room,li2015observation}, valley-selective transport behaviors \cite{belayadi2023valley,torres2019valley,zhai2018local,li2023strain}, and strain-induced quantum interference \cite{myoung2020manipulation}. Moreover, nanobubbles can host localized states that exhibit conductance resonances, indicating that the presence of nanobubbles significantly influences the detailed structure of the density of states (DOS) \cite{myoung2020manipulation}. Although the DOS provides comprehensive information about the system, directly extracting specific physical details about nanobubbles from the DOS is challenging. This difficulty arises from the simultaneous presence of both local and global features within the aggregated data.

\begin{table*}[tbp]
\caption{\label{tb.labels_features} The superscripts A, B, and C denote indices of single-bubble data that are combined to generate multibubble samples, with indices satisfying $0 \leq A < B < C \leq 274$ to ensure no duplication of samples. For the dataset accommodating up to $N_\mathrm{bubble}$ bubbles, the labeling system comprises $2N_\mathrm{bubble} + 1$ labels for each sample: one discrete label indicating the actual number of nanobubbles, and $2N_\mathrm{bubble}^\mathrm{max}$ continuous labels representing the sizes of the bubbles. The value of the discrete label is determined by rounding the machine learning prediction to the nearest integer. The DOS spectra, \D, are calculated by averaging the interpolated DOS spectra, $D(E)_{\mathrm{int}}$, followed by taking the natural logarithm of this average.}
    \begin{tabularx}{\textwidth}{>{\hsize=.3\hsize}X>{\hsize=1.2\hsize}X>{\hsize=1.5\hsize}X}

\hline
\Nbubble & Labels &   Features (DOS)\\
  \hline
 0 &  $[0, 0, 0, 0, 0, 0, 0]$ & $\log[D(E)_{\mathrm{int}}^{(Z)}]$\\
 1 &  $[1, \sigma^{(A)}, h_{0}^{(A)}, \sigma^{(A)}, h_{0}^{(A)}, \sigma^{(A)}, h_{0}^{(A)}]$ & $\log[D(E)_{\mathrm{int}}^{(A)}]$\\
 2 &  $[2, \sigma^{(A)}, h_{0}^{(A)}, \sigma^{(B)}, h_{0}^{(B)}, \sigma^{(B)}, h_{0}^{(B)}]$ & $\log[ (D(E)_{\mathrm{int}}^{(A)} + D(E)_{\mathrm{int}}^{(B)}) / 2]$\\
 3 &  $[3, \sigma^{(A)}, h_{0}^{(A)}, \sigma^{(B)}, h_{0}^{(B)}, \sigma^{(C)}, h_{0}^{(C)}]$ & $\log[ (D(E)_{\mathrm{int}}^{(A)} + D(E)_{\mathrm{int}}^{(B)} + D(E)_{\mathrm{int}}^{(C)}) / 3]$\\
  \hline
\end{tabularx}
\end{table*}

This study introduces a novel strategy for detecting and identifying nanobubbles in monolayer graphene by leveraging machine learning (ML) techniques to navigate the complexities inherent in processing data from electrical measurements. Building on prior research in nanobubble identification \cite{song2021machine,wu2022rapid} and the advancement of a graphene nanobubble sensor using quantum interferometry \cite{myoung2023detecting}, our approach emphasizes the identification of multiple nanobubbles. It aims to develop an accurate predictive model that directly correlates the geometric characteristics of nanobubbles with the DOS in graphene samples ~\cite{fung2021dosnet}. This direct correlation enables the precise assignment of measurements to individual nanobubbles within a sample, even in scenarios involving multiple nanobubbles. Given that nanobubbles can significantly alter the electrical properties of graphene, particularly in cases of degenerate defect states, the exploration of an ML approach to recognize multiple nanobubbles is deemed essential. Our investigation into the effects of nanobubble multiplicity on the ML recognition algorithm reveals an effective technique for accurately identifying individual nanobubbles. The proposed ML-based method for nanobubble recognition is poised to enhance the pure-electrical evaluation of graphene and other 2D materials, potentially streamlining the mass production of these materials for device fabrication.


\section{Methods}
\subsection{Generating Raw Nanobubble Data}
We approximated the nanobubble using a Gaussian distribution located at $\vec{r}_{0}$ as
\begin{align}
    h( \vec{r} ) = h_{0}e^{-(\vec{r}-\vec{r} _{0})^{2}/2\sigma^{2}},
\end{align}
where \h is the maximum height of the vertical deformation and \s is the standard deviation corresponding to the width of the nanobubble.
This nanobubble introduces a PMF, which can be described by the same mechanism as electron confinement in real magnetic fields \cite{peres2006dirac,tan2010graphene,myoung2019splitting,mills2019dirac}. The $C_{3v}$ symmetric PMF is described by
\begin{align}
    \vec{B}_{\text{ps}}( \vec{r} ) = \nu \cfrac{\hbar \beta}{e a_{0}} \cfrac{h_{0}^{2}}{\sigma^{6}} r^{3} e^{-(r-r_{0})^{2}/2\sigma^{2}} \sin 3\theta \hat{z},
\end{align}
where $\beta=3.37$, $a_{0}=0.146~\mathrm{nm}$ is the nearest carbon-carbon bonding length, and $\nu=\pm1$ for different valleys of graphene, respectively.
We set its maximum strength $\tilde{B}_{\text{max}}$ at $r-r_{0}=\sqrt{3/2}\sigma$ as the representative quantity of nanobubbles, which is formulated as 
\begin{align}
    \tilde{B}_{\text{max}}(\sigma, h_{0}) \equiv \cfrac{h_{0}^{2}}{\sigma^{3}}.
\end{align}

In the presence of nanobubbles, DOS spectra exhibit peaks at specific energies, resulting from PMF-induced localized states within nanobubbles. As \B increases, a greater number of localized states emerge, and their corresponding DOS peaks tend to shift toward lower energy. This correlation between DOS and \B is a primary focus of this work, analyzed through a supervised machine learning approach. As a theoretical study, we obtained the DOS spectra through tight-binding calculations of a graphene sheet with nanobubbles, employing \textsc{kwant} codes~\cite{groth2014kwant}. While the number of bubbles could be further extended, our study focuses on $N_{\text{max}} = 3$ case, indicating that the maximum number of bubbles considered is 3. This consideration is justified as the standard preprocessing involving heating and cooling of the graphene leads to the coalescence of adjacent nanobubbles, thereby limiting the total number of bubbles within a single sheet.

\subsection{Preprocessing Nanobubble Data}
The raw DOS spectra consist of 20,001 data points spanning an energy range from -0.1 to 0.1 eV; subsequently, we performed interpolation on the DOS to standardize the energy grid, resulting in a range from -0.099 to 0.099 eV with 10,001 data points ~\cite{Virtanen2020scipy}. Each DOS spectrum, $D(E)_{\mathrm{int}}$, is characterized by two important parameters, $\sigma$ and $h$, corresponding to the geometrical factors of nanobubbles, as previously mentioned. We considered $\sigma$ ranging from 20 to 30 $a$ and $h$ from 1 to 25$a$, with a step of $a$, where $a = \sqrt{3}a_{0} = 0.246 \mathrm{nm}$ is the lattice constant of graphene. Consequently, the total dataset comprises 275 samples (= 11 $\times$ 25), marking a significant expansion compared to previous research~\cite{song2021machine}, which utilized 75 samples.

In this work, we generate a multibubble dataset by combining data from the single-bubble dataset. The multibubble DOS is equivalent to the average of the corresponding single-bubble DOS data, within numerical accuracy, provided that the bubbles are sufficiently distant from each other. If the bubbles are spatially close to each other, they merge into a single bubble for stabilization. Therefore, for our analysis, we can simply aggregate single-bubble DOS data to generate a multibubble dataset.

For $N_{\mathrm{bubble}} = 1, 2$, we include all 275 single-bubble samples and all possible combinations between them, totaling $275 \times 274 / 2 = 37,675$. For $N_{\mathrm{bubble}} = 3$, the number of available combinations exceeds our computational capacity, so we randomly select 37,675 samples from the three-bubble combinations. Additionally, we include a dataset where $N_{\mathrm{bubble}} = 0$, as a reference to determine whether a given DOS contains nanobubbles or not.

We introduce a unified labeling scheme for varying numbers of nanobubbles, as illustrated in Table~\ref{tb.labels_features}. In practice, the accidental occurrence of two nanobubbles of identical size on a single graphene sheet is extremely rare. Therefore, we treat a set of nanobubbles with the same index as a single nanobubble.

We divided the dataset into training and testing sets with a train-test ratio of 7:3 for each $N_{\mathrm{bubble}}$ case to mitigate the risk of an imbalanced dataset, which could adversely affect the model's performance. To prevent overfitting, we employed data augmentation techniques, such as min-max scaling and the addition of appropriate white noise. This ensured that each $N_{\mathrm{bubble}}$ occupies the same partition in the training and test set, resulting in a total sample ratio of approximately 7:3 (527,100 samples in the training set and 226,400 samples in the test set). The DOS spectra, $\tilde{D}(E)$, serving as features for the supervised machine learning, are defined as follows ~\cite{buitinck2013api}:
\begin{align}
\tilde{D}(E) = \frac{D(E) - \min(D(E))}{\max(D(E)) - \min(D(E))} + \eta\xi(E)
\end{align}
where $\eta$ denotes the amplitude of noise, and $\xi(E)$ represents a random variate sampled from a normal distribution $\mathcal{N}(0,1)$. By adjusting $\eta$, we introduce controlled measurement noise into the dataset.

\subsection{Deep learning model parameters}
\begin{figure}[tbp]
    \centering    \includegraphics[width=0.6\columnwidth]{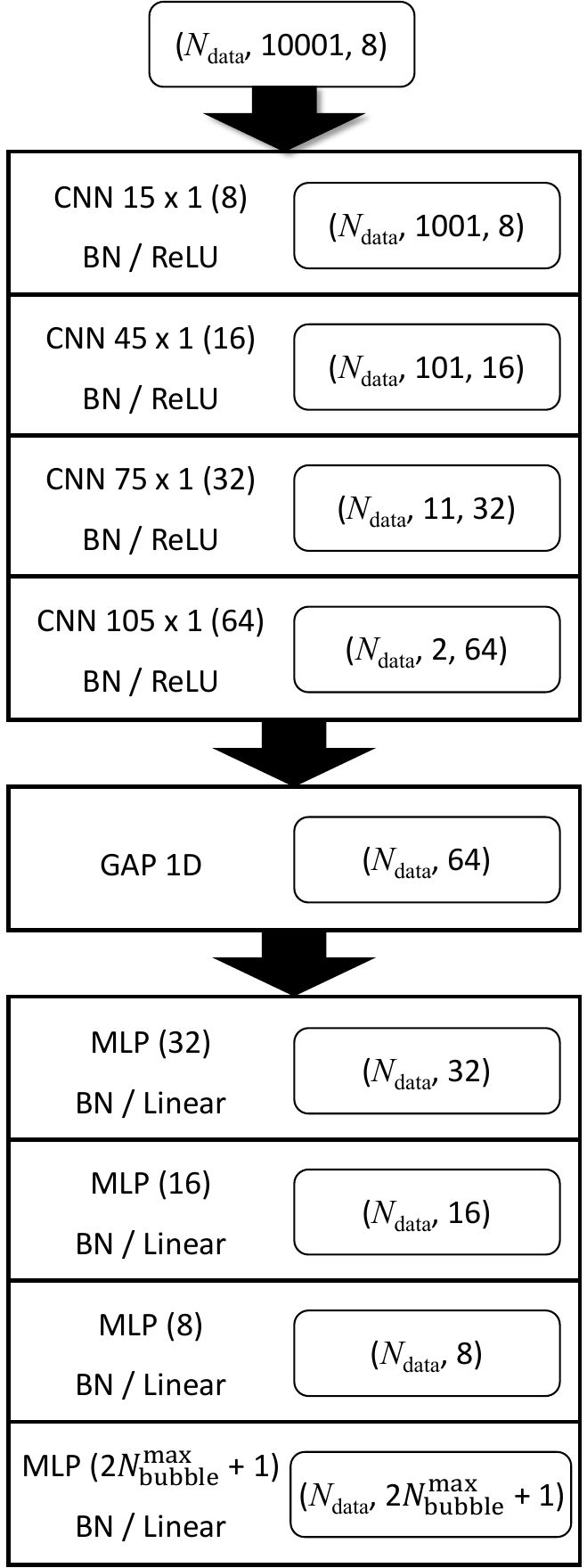}
    \caption[]{Deep learning model architecture for recognition of nanobubbles.
    The architecture of the Deep Learning Model for Nanobubble Recognition. 
    The abbreviations `CNN, `GAP 1D', and `MLP' refer to the Convolutional Neural Network, One-Dimensional Global Average Pooling, and Multi-Layer Perceptron, respectively. 
    `BN', `ReLU', and `Linear' correspond to the Batch-Normalization process, Rectified Linear Unit activation function, and Linear activation function, respectively. 
    In each CNN layer, the kernel size and number of filters are specified adjacent to CNN, with all layers sharing a uniform stride of 10. For the MLP layers, unit parameters are detailed next to MLP.
     The output tensor shapes for each layer are depicted within squares featuring rounded corners. $N_{\mathrm{data}}$ and \Nmax denote the dataset size and the maximum number of nanobubbles, respectively. }
    \label{fig:model_architecture}
\end{figure}

Figure~\ref{fig:model_architecture} illustrates the architecture of our deep learning model. As outlined in Table~\ref{tb.labels_features}, we utilize a discrete label and $2N_{\mathrm{bubble}}$ continuous labels to represent the number of nanobubbles and the geometrical shapes of the nanobubbles, respectively.
The model serves dual functions: it operates as a classifier using the discrete label and as a regressor for the continuous labels.

The deep learning model is developed using TensorFlow packages and consists of three primary components: a 1D Convolutional Neural Network (CNN), 1D Global Average Pooling (GAP), and a Multi-Layer Perceptron (MLP)\cite{tensorflow2015-whitepaper}. Linear and Rectified Linear Unit (ReLU) activation functions are employed within our model. The CNN layers are organized into blocks as follows: [1D CNN layer - Batch Normalization (BN) \cite{pmlr-v37-ioffe15} - ReLU], with filter sizes of [8 – 16 – 32 – 64], kernel sizes of [15 – 45 – 75 – 105], and all strides set to 10. This arrangement of CNN layers facilitates the extraction of the feature tensor from the input data. Subsequently, a 1D GAP layer is used to flatten the feature tensor, preparing it for the MLP component, which is composed of blocks structured as [MLP layer - BN - Linear]. The unit sizes in the MLP layers are determined to be [32 – 16 – 8 – (2\Nmax+1)], where \Nmax denotes the maximum number of nanobubbles, set at 3 in this instance.

To facilitate effective training, we employ the Adaptive Moment Estimation (Adam) optimizer~\cite{kingma2017adam} along with a Mean Squared Error (MSE) loss function. For validation, 20\% of the training set is randomly selected and shuffled at the beginning of each epoch. Utilizing callback functions from TensorFlow, we monitor the validation loss value for each epoch and optimize our model by aiming to achieve the minimum validation loss.


\begin{figure*}[btp]
    \centering
    \includegraphics[width=\textwidth]{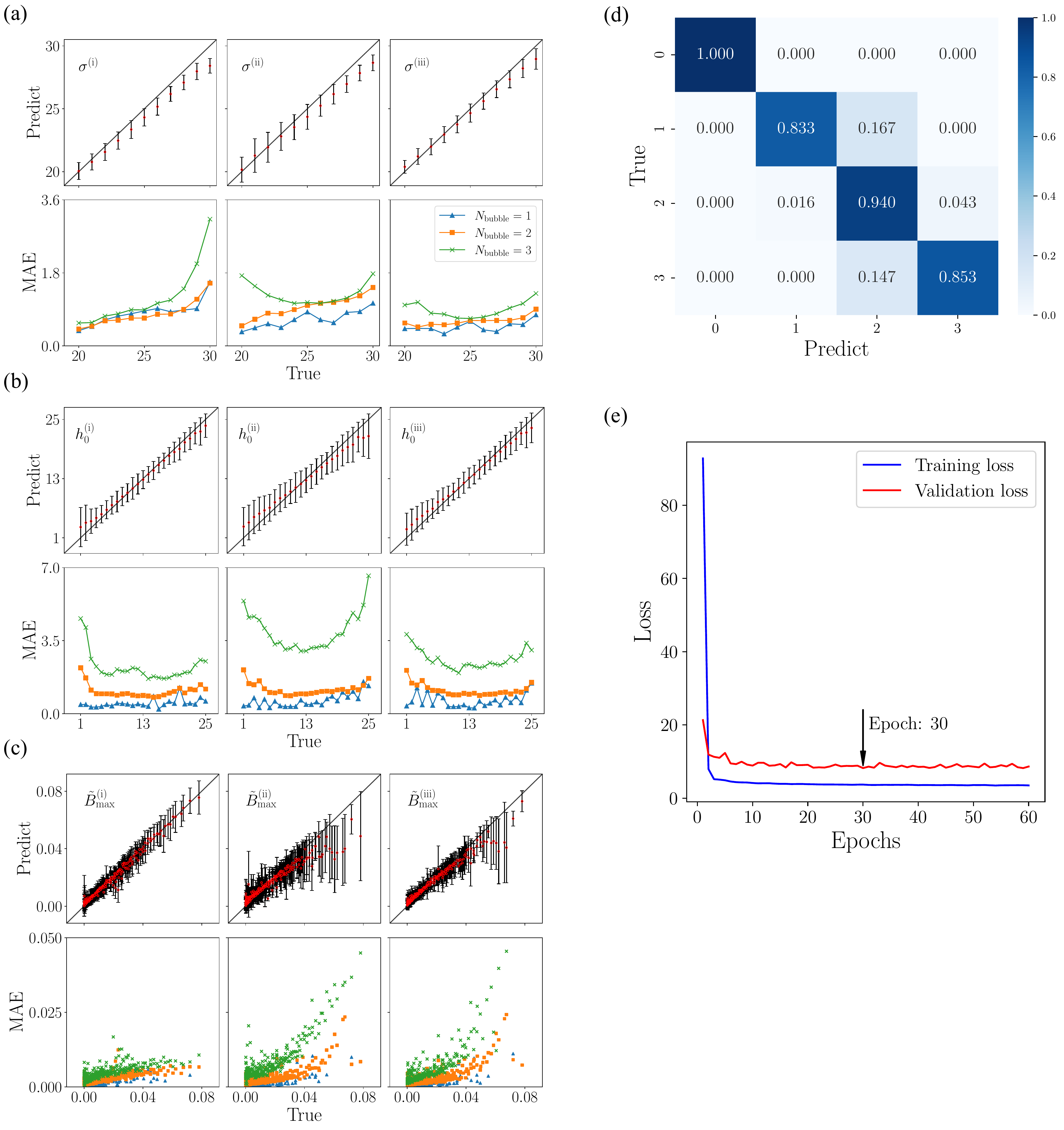}
    \caption[]{Recognition performance of our model upon the \D spectrum datasets. (a)-(c)  Regression results of \s, \h, and \B, respectively. The top panels indicate the predicted values of individual nanobubbles when \Nmax=3, and the bottom panels show the prediction errors for different \Nbubble. The diagonal lines in the top panels indicate the exact prediction lines. (d) Confusion matrix for classification results of the datasets with different \Nbubble. (e) Training and validation loss as a function of epoch. The optimal epoch is marked by a black arrow.}
    \label{fig2}
\end{figure*}

\section{Results and Discussion}

Figure~\ref{fig2} showcases the performance results for both regression and classification. The regression analysis evaluates the precision of predicted values for \s, \h, and \B of individual nanobubbles, compared to their true values, denoted by superscripts $\mathrm{(i)}$, $\mathrm{(ii)}$, and $\mathrm{(iii)}$ respectively.

In Figure~\ref{fig2} panels (a-c), the regression outcomes for \s, \h, and \B are presented. Overall, the mean values and standard deviations closely align with the diagonal lines, illustrating the model's accurate prediction of nanobubble parameters. Additionally, the precision of prediction performance is further assessed by calculating the mean absolute errors (MAE) for the nanobubble parameters \s, \h, and \B for individual bubbles,
\begin{align}
\mathrm{MAE} = \cfrac{1}{N} \sum_{i=1}^N |x_i - \hat{x}_{i}|,
\end{align}
where $N$ denotes the total number of datasets, $x_{i}$ represents the true value, and $\hat{x}_{i}$ is the predicted value. As demonstrated in Figure~\ref{fig2}(a-c), the prediction errors for the nanobubble parameters when $N_{\mathrm{bubble}}=1$ are found to be minimal, attributed to the uniformity of labels (i.e., [1, A, A, A]). Conversely, for $N_{\mathrm{bubble}}>1$, the labels within the dataset are not uniform but distinct, leading to larger MAEs.

In addition to \s and \h, we also analyze the performance of our deep learning model based on the parameter \B. In the classification of the DOS spectra, we find that \B serves as the representative parameter, rather than  \s and \h. Given that the electronic states of strained graphene are determined by characteristic PMF profiles, it is logical to predict that graphene nanobubbles with similar \B values will produce analogous DOS spectra, despite variations in \s and \h values.

Figure~\ref{fig2}(c) illustrates the prediction accuracy of such a representative physical quantity, \B, which we calculated using predicted \s and \h values. The results show a good agreement between the predicted and true \B values, surpassing the prediction accuracy for \s and \h. Additionally, it is observed that the prediction accuracy slightly declines for larger \B values. Nanobubbles corresponding to these larger \B cases are generally characterized by a high aspect ratio of \s and \h, that is, smaller \s and larger \h. From a theoretical perspective, such high \B nanobubbles are likely to result in significant errors, as nanobubbles with small \s cannot be accurately modeled by elastic theory under the assumption of a continuous graphene sheet. Therefore, we focus on moderate or small \B cases, where the accuracy of our machine learning recognition remains notably high.

\begin{table}[tbp]
\caption{Classification performance of \Nmax = 3 model according to evaluation metrics, where the accuracy scores 0.906.}
\renewcommand*{\arraystretch}{1.4}
\begin{center}
    \begin{tabularx}{0.95\columnwidth}{XXX}
  \hline
 \Nbubble& Precision&Recall\\
 \hline
 0 & 1.000& 1.000\\
 1 & 0.981& 0.833\\
 2 & 0.749& 0.940\\
 3 & 0.951& 0.853\\
  \hline
\end{tabularx}
\label{classification-table}
\end{center}
\end{table}

The classification performance of \Nbubble is evaluated using a confusion matrix. As displayed in Figure~\ref{fig2}(d), the confusion matrix from the classification results clearly demonstrates that our model effectively recognizes the number of nanobubbles present in a graphene sample by analyzing given DOS spectra. For this confusion matrix, we employ the conventional evaluation metrics as follows ~\cite{SOKOLOVA2009427}:
\begin{align}
A = \cfrac{\Sigma_{i}M_{ii}}{\Sigma_{i}\Sigma_{j}M_{ij}},\ 
P_{i} = \cfrac{M_{ii}}{\Sigma_{j}M_{ji}},\ 
R_{i} = \cfrac{M_{ii}}{\Sigma_{j}M_{ij}},
\end{align}
where $A$ represents accuracy, $P_{i}$ denotes precision, and $R_{i}$ signifies recall for each \Nbubble = $i$ ($0 \leq i \leq 3$). Precision measures the proportion of retrieved values that are relevant, while recall quantifies the proportion of relevant values that are successfully retrieved. Table~\ref{classification-table} showcases the computed evaluation metrics for our model. Our model demonstrates superior performance with smaller $N_\mathrm{bubble}$, yet remains effective even with larger numbers of bubbles. The optimal model performance is obtained at the 30th epoch, as shown in Figure~\ref{fig2}(e). The gap between the training and validation loss does not fluctuate heavily but shows stable convergence. Furthermore, we can systematically improve the performance by providing additional training samples for a larger number of bubbles.
\begin{figure}[tbp]
    \centering
    \includegraphics[width=\columnwidth]{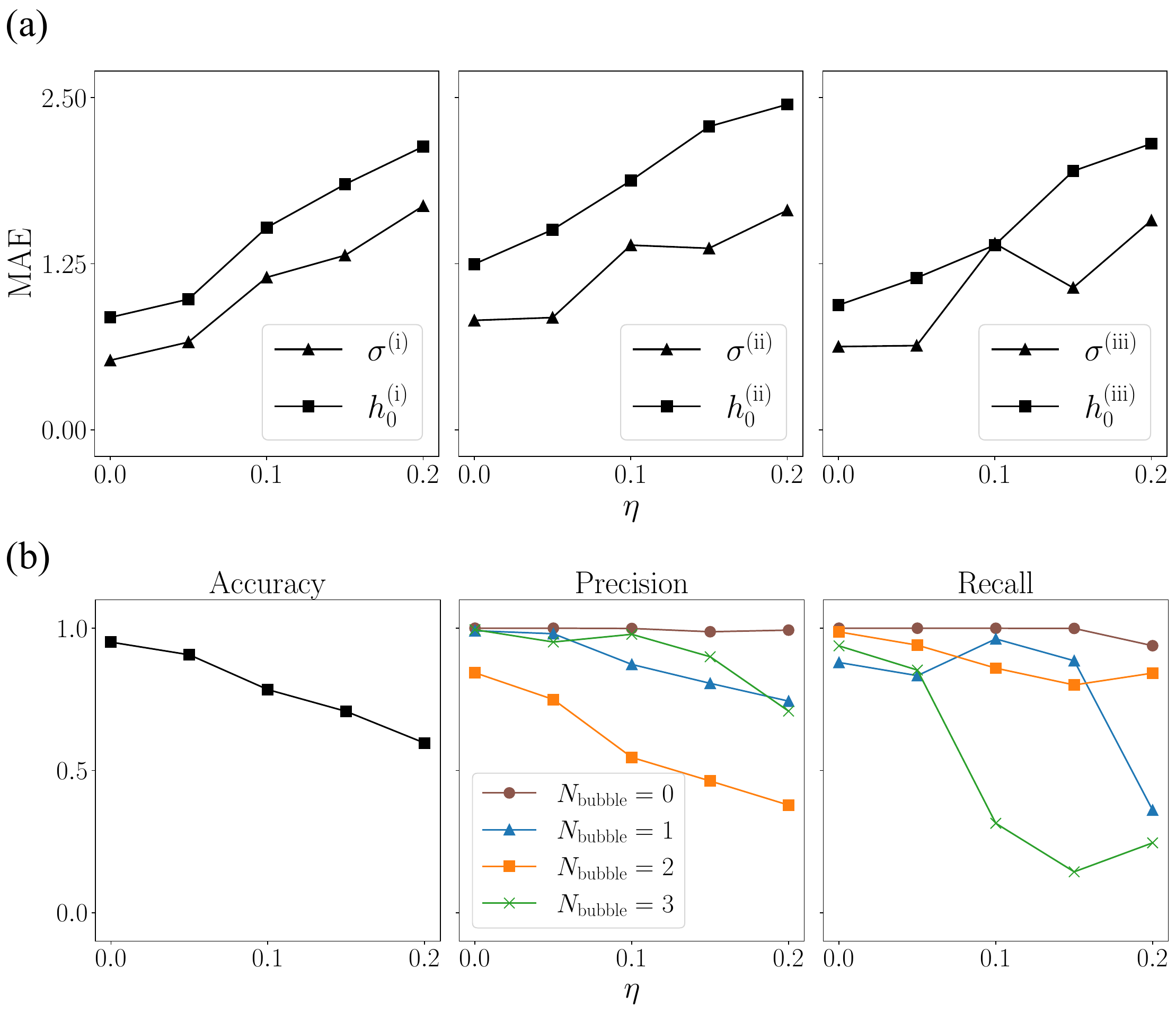}
    \caption[]{Recognition performance with respect to noise amplitude \e. 
    (a) Regression results of \s and \h are depicted by MAE for each model with varying \e. Prediction errors for both \s and \h increases as \e increases. 
    (b) Classification results are presented using evaluation metrics. The accuracy derived from the confusion matrix decreases as \e increases.}
    \label{fig3}
\end{figure}

In practical experimental situations, we expect the DOS signals to contain random noise from measurements. Therefore, the model's performance in the presence of noise is a key component for applying the model to real-world samples.
We examine the effects of noise on our model's prediction and classification performance. We introduce white noise by adding a uniform random distribution to the DOS spectra, with an amplitude $\eta$. Figure~\ref{fig3} demonstrates that both the regression and classification performance deteriorate as $\eta$ increases. The analysis shows that the lower the noise level $\eta$, the better the model performs, as expected. While solely based on this noise level dependency, it might seem that training a model with higher noise strength is not beneficial, cross-noise level testing reveals different conclusions.

\begin{figure}[tbp]
    \centering
    \includegraphics[width=\columnwidth]{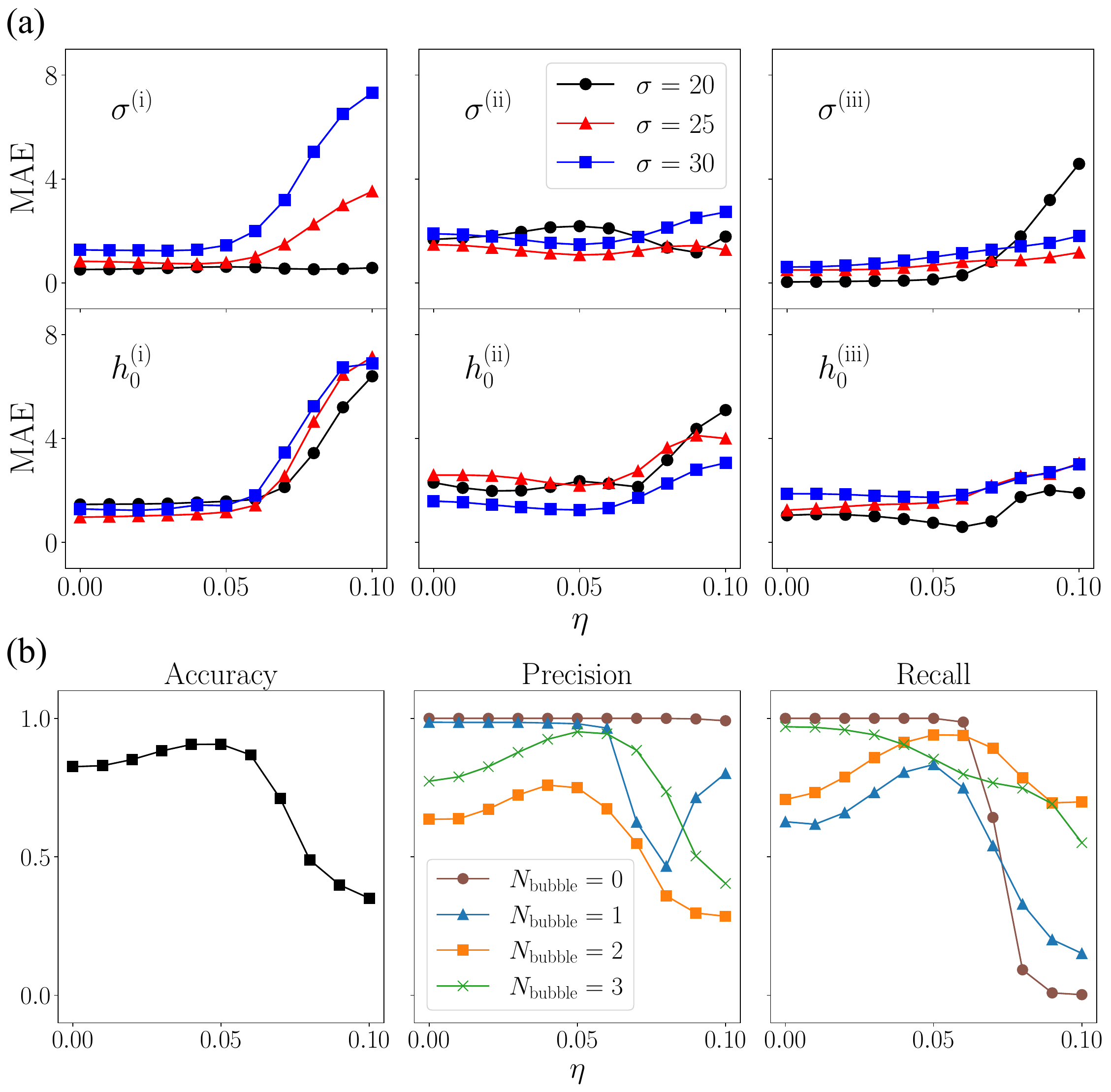}
    \caption[]{Performance of the trained model with \e = 0.05 when subjected to test data \D with varying values of \e. 
    (a) Regression results are illustrated through MAE with the selected dataset for \h = 13 and \s = 20, 25, 30. 
    (b) Classification performances are presented using evaluation metrics.}
    \label{fig4}
\end{figure}

After training a model with the $\eta=0.05$, we apply the model to the DOS dataset with different noise levels. To visualize the recognition performance for different noise levels, we choose specific samples $h_{0}=13$ and $\sigma=20,25,30$, but the overall tendency is independent of the choice of the samples. The regression and classification evaluations are carried out with the same \D dataset by varying the noise level $0\le \eta\le 0.1$. As shown in Figure~\ref{fig4}(a), it is evident that our model exhibits good performance when the noise amplitudes for the evaluation are equal to or less than the trained model's $\eta$. Interestingly, the prediction performance does not deteriorate much even if the noise level is slightly higher than the trained $\eta=0.05$.

Similarly, as shown in Figure~\ref{fig4}(b), the classification performance of our model exhibits high accuracy when the noise levels of the evaluation datasets do not exceed that of the trained dataset. A notable aspect of the classification results is the optimization of model performance at noise levels in the evaluation dataset that match the trained dataset's noise level. This observation suggests that our deep learning model is adept at effectively detecting nanobubbles in \D when operating within an appropriate noise amplitude, $\eta$, consistent with the model's training conditions. Consequently, it is crucial to choose an appropriate noise amplitude during training, considering an optimal range of $\eta$ to ensure high accuracy.


\section{Results of \Nmax = 5 model}


\begin{figure}[tbp]
    \centering
    \includegraphics[width=0.8\columnwidth]{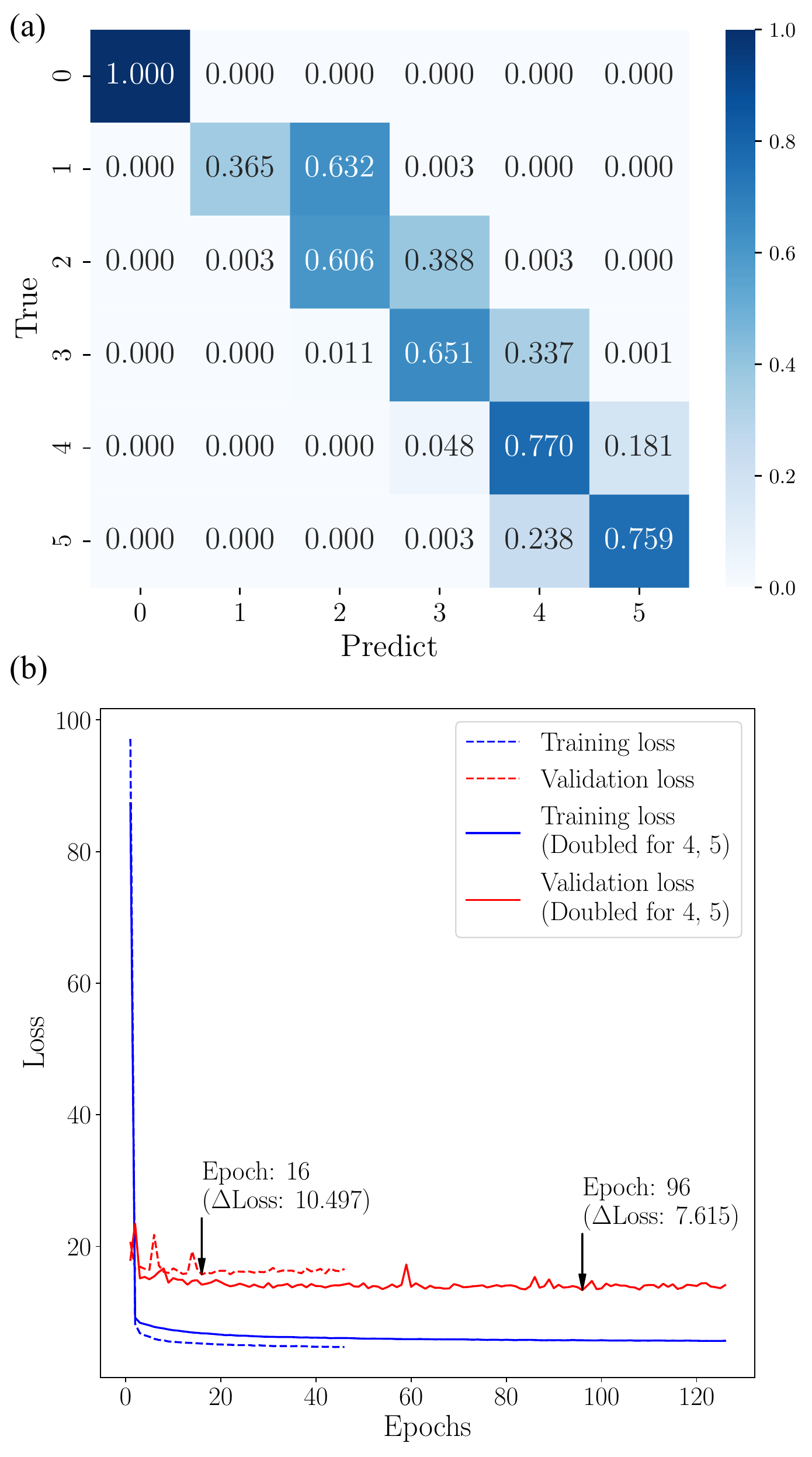}
    \caption[]{(a) Confusion matrix for the \Nmax = 5 model applied to a doubled dataset for \Nbubble = 4 and 5, while maintaining the same size for the other datasets. (b) Learning curves comparison for the \Nmax = 5 model. The doubled dataset for \Nbubble = 4 and 5 decreases the loss gap. }
    \label{fig5}
\end{figure}

For our next step, we aim to generalize our recognition model to handle larger \Nmax cases beyond 3. To achieve this, we apply the same methodology to a dataset with \Nmax = 5, using the same algorithm as described in the previous section. With \Nmax$=5$, the number of possible combinations from the dataset exceeds the predefined maximal combination $_{275}C_2$. Consequently, we augment the dataset to ensure that the maximum number of combinations is effectively doubled for $N_\mathrm{bubble}=4$ and 5, compensating for the relatively small portion of all possible combinations included in the training set.


With the augmented dataset, we assess the regression performance, finding an accuracy of 0.71. While the accuracy of our model for \Nmax$=5$ may not be significantly higher compared to the \Nmax$=3$ scenarios, this moderate level of accuracy could still be meaningful for generalization purposes. The classification performance for \Nmax$=5$ with the augmented dataset is illustrated in Figure~\ref{fig5}. Our model performs effectively for a larger number of cases, yet exhibits lower performance for scenarios with fewer cases. The decrease in classification performance for fewer \Nbubble cases is attributed to the limited variety within the dataset for \Nbubble~=~1 and \Nbubble~=~2, even after dataset augmentation. The difference between the training and validation loss is relatively large in comparison to the \Nmax$=3$ case but data augmentation reduces the gap. As shown by the decreased loss gap with additional samples in Figure~\ref{fig5}, providing sufficiently large datasets for training could further enhance our model’s ability to effectively recognize \Nbubble. In essence, our deep learning approach remains valid when \Nbubble is less than or equal to the \Nmax of the training dataset.

\section{Summary}
We have explored the recognition of multiple nanobubbles in graphene using DOS spectra through neural network techniques, aiming to expand upon the previously demonstrated single-bubble recognition \cite{song2021machine}. To facilitate the recognition of multiple nanobubbles, we devised a labeling rule based on the premise that multiple nanobubbles with identical indices can be treated as a single bubble, given their low likelihood of simultaneous occurrence in practical scenarios.

Our study demonstrates commendable performance in both regression and classification for a model with \Nmax = 3. Notably, the trained model shows high proficiency in distinguishing whether a given DOS spectrum represents graphene with or without nanobubbles. Moreover, the model maintains effective recognition capabilities for nanobubbles when the DOS spectra are subjected to noise levels with an amplitude that is equal to or less than the model's trained noise level.

We also attempted to extend our model to accommodate larger \Nmax values to generalize its application. Although the model's performance diminishes slightly in comparison to the \Nmax$=3$ scenario, achieving a moderate level of accuracy, we identified that the diminished performance for larger \Nmax values could be mitigated by increasing the dataset size, thus enabling significant enhancements in nanobubble recognition accuracy. Consequently, by incorporating additional training datasets into the ML model, our approach can be adapted to manage larger \Nmax values effectively.

\section*{Acknowledgments}
SK and AG are supported by the National Research Foundation under grant numbers NRF-2021R1C1C1010429 and NRF-2023M3K5A1094813. NM and SJ acknowledge the support of the National Research Foundation through the Korean governments NRF-2022R1F1A1065365 (MIST) and RS-2023-00285353 (MOE).


\bibliographystyle{elsarticle-num-names}
\bibliography{MLBubbleSLG}

\end{document}